\documentclass[a4paper,fleqn,sort&compress]{cas-dc}

\usepackage{amsmath}
\DeclareMathOperator\arctanh{arctanh}
\usepackage{cite}
\usepackage[numbers]{natbib}
\usepackage{balance}
\usepackage{caption}
\usepackage{soul,color,xcolor}
\def\tsc#1{\csdef{#1}{\textsc{\lowercase{#1}}\xspace}}
\tsc{WGM}
\tsc{QE}

\begin{document}

\let\WriteBookmarks\relax
\def\floatpagepagefraction{1}
\def\textpagefraction{.001}
\let\printorcid\relax

\shorttitle{A dynamic thermal sensing mechanism with reconfigurable expanded-plane structures}    

\shortauthors{H. Tan, H. Cai and P. Jin et al.}  

\title [mode = title]{A dynamic thermal sensing mechanism with reconfigurable expanded-plane structures}  



%

\author[1]{{Haohan Tan}}





\credit{Formal analysis, Writing - orginal draft}

\affiliation[1]{organization={Department of Physics, State Key Laboratory of Surface Physics, and Key Laboratory of Micro and Nano Photonic Structures (MOE), Fudan University},
            city={Shanghai},
            postcode={200438}, 
            country={China}}

\author[1]{{Haoyang Cai}}
\credit{Formal analysis, Writing - orginal draft}

\author[1]{{Peng Jin}}




\credit{Conceptualization, Writing - orginal draft}
\cormark[1]

\ead{19110190022@fudan.edu.cn}

\author[1]{{Jiping Huang}}
\credit{Conceptualization, Writing - orginal draft}
\cormark[1]

\cortext[1]{Corresponding author.}

\ead{jphuang@fudan.edu.cn}



\begin{abstract}
  The precise measurement of temperature is crucial in various fields such as biology, medicine, industrial automation, energy management, and daily life applications. While in most scenarios, sensors with a fixed thermal conductivity inevitably mismatch the analogous parameter of the medium being measured, thus causing the distortion and inaccurate detection of original temperature fields. Despite recent efforts on addressing the parameter-mismatch issue, all current solutions are constrained to a fixed working medium whereas a more universal sensor should function in a variety of scenes. Here, we report a dynamic thermal sensor capable of highly accurate measurements in diverse working environments. Remarkably, thanks to the highly tunable thermal conductivity of the expanded-plane structure, this sensor works effect on background mediums with a wide range of conductivity. Such a development greatly enhances the robustness and adaptability of thermal sensors, setting a solid foundation for applications in multi-physical sensing scenarios.
\end{abstract}




\maketitle

\section{Introduction}\label{Introduction}

{Over the past few decades, the field of thermal metamaterials has witnessed significant advancements~\cite{fan2008shaped, xu2023giant, xu2022thermal, zhou2023adaptive, yangmass, jin2023deep}. Researchers have introduced an array of devices with novel functionalities, encompassing thermal cloaks~\cite{yang2021optimization, shen2016thermal, han2023itr, fujii2018exploring}, concentrators~\cite{moccia2014independent, chen2015experimental, xu2019converging, xu2018thermal, ji2021designing, xu2019bilayer, fujii2020cloaking, chen2015materials}, rotators~\cite{yang2020experimental, guenneau2013anisotropic}, illusion~\cite{zhu2015converting,yang2019thermal}, transparency~\cite{xu2019thermal}, and sensors based on thermotics and scattering cancellation techniques~\cite{xu2023transformation, li2021transforming, yang2016thermal, guenneau2012transformation}. Notably, the concept of topology has also been integrated into thermal metamaterials~\cite{sha2022topology1, xu2023topology, jin2023tunable, sha2022topology, hostos2023design}. Among these devices, thermal sensors have garnered considerable attention due to their pivotal role in temperature field detection. However, a prevalent shortcoming of conventional thermal sensors is their proclivity to distort the measured temperature field, resulting in inaccuracies in the obtained results. To address this issue, various solutions have been proposed~\cite{xu2020thermally, wang2021multithermally, jin2020making, jin2021particle}. For instance, Xu et al. introduced the concept of a thermal invisible sensor~\cite{jin2020making, xu2020thermally}. By solving the linear Laplace equation, they derived two groups of thermal conductivities to realize thermally invisible sensors, encompassing geometrically anisotropic cases. Their bilayer scheme can maintain the original temperatures in both the sensor and the background. Subsequently, considering the combination of radiation and conduction, described by Fourier's law and Rosseland diffusion approximation, respectively, Wang et al. accommodated multiphysics fields~\cite{wang2021multithermally} to achieve thermal sensors. Jin et al. also proposed an anisotropic monolayer scheme to prevent thermal sensors from distorting local and background temperature profiles, thereby ensuring accuracy and thermal invisibility~\cite{jin2020making}. They further introduced an optimization model utilizing particle swarm algorithms to create bilayer thermal sensors composed of bulk isotropic materials~\cite{jin2021particle, zhang2022extracting}. They selected suitable materials for different regions and treated the radii of the sensor, inner shell, and outer shell as design variables, ultimately minimizing the fitness function through particle swarm optimization. Their model can also be flexibly extended to the square case.

Nevertheless, despite substantial efforts to address the parameter-mismatch issue, current solutions primarily concentrate on a fixed background environment where the thermal conductivity of the background region remains constant. This raises a crucial unresolved problem: how to design a versatile and reconfigurable sensor capable of adapting to varying background environments. In this context, "reconfigurability" refers to the sensor's ability to effectively operate amid changes in the background environment. Existing studies indicate that, for bilayer sensors, any alteration in the thermal conductivity of the background region necessitates adjustments in the thermal conductivity of the core or shell region. In practical applications, this would entail altering the material composition of the core or shell region, which becomes impractical due to the wide range of potential background environments. Recently, an innovative type of thermal metamaterials based on an expanded-plane (EP) configuration has been proposed~\cite{guo2022passive}. This structure not only facilitates ultra-high effective thermal conductivity but also allows for the adjustment of effective thermal conductivity by simply modifying the height of the expanded plane, thereby offering new possibilities for the development of reconfigurable thermal sensors (refer to Fig. 1(a)-(d)). Building upon the EP structure, Han et al. have also designed a thermal cloak using a single material, thus eliminating interfacial thermal resistance~\cite{han2023itr, nguyen2015active, yue2021thermal, han2013homogeneous, xu2014ultrathin}.

In this paper, we present the design of a single expanded-plane (SEP) bilayer sensor and evaluate its performance through both simulations and experiments. Among various numerical methods~\cite{chaurasiya2023numerical, chaurasiya2023taylor, chaurasiya2023numerical1, chaurasiya2023temperature, chaurasiya2022study}, we opt for the finite-element simulation method due to its distinct advantages. The results unequivocally affirm that by simply adjusting the height of the expanded plane, the SEP sensor can efficiently operate in a diverse range of environments, thereby demonstrating its reconfigurability. Furthermore, similar outcomes are obtained when employing multiple expanded-planes (MEP) structures.

In comparison to existing research, our design demonstrates adaptability to environments characterized by varying temperature differences between hot and cold sources. The EP structure also exhibits excellent performance in the presence of point heat sources. This adaptability is realized through the addition or removal of copper rings on the EP structure. These findings not only broaden the potential applications of thermal sensors but also hold significant implications for the wider field of thermal detection, including industrial applications, environmental monitoring, and medical applications.}
\begin{figure*}[htp]
  \includegraphics[width=1.0\linewidth]{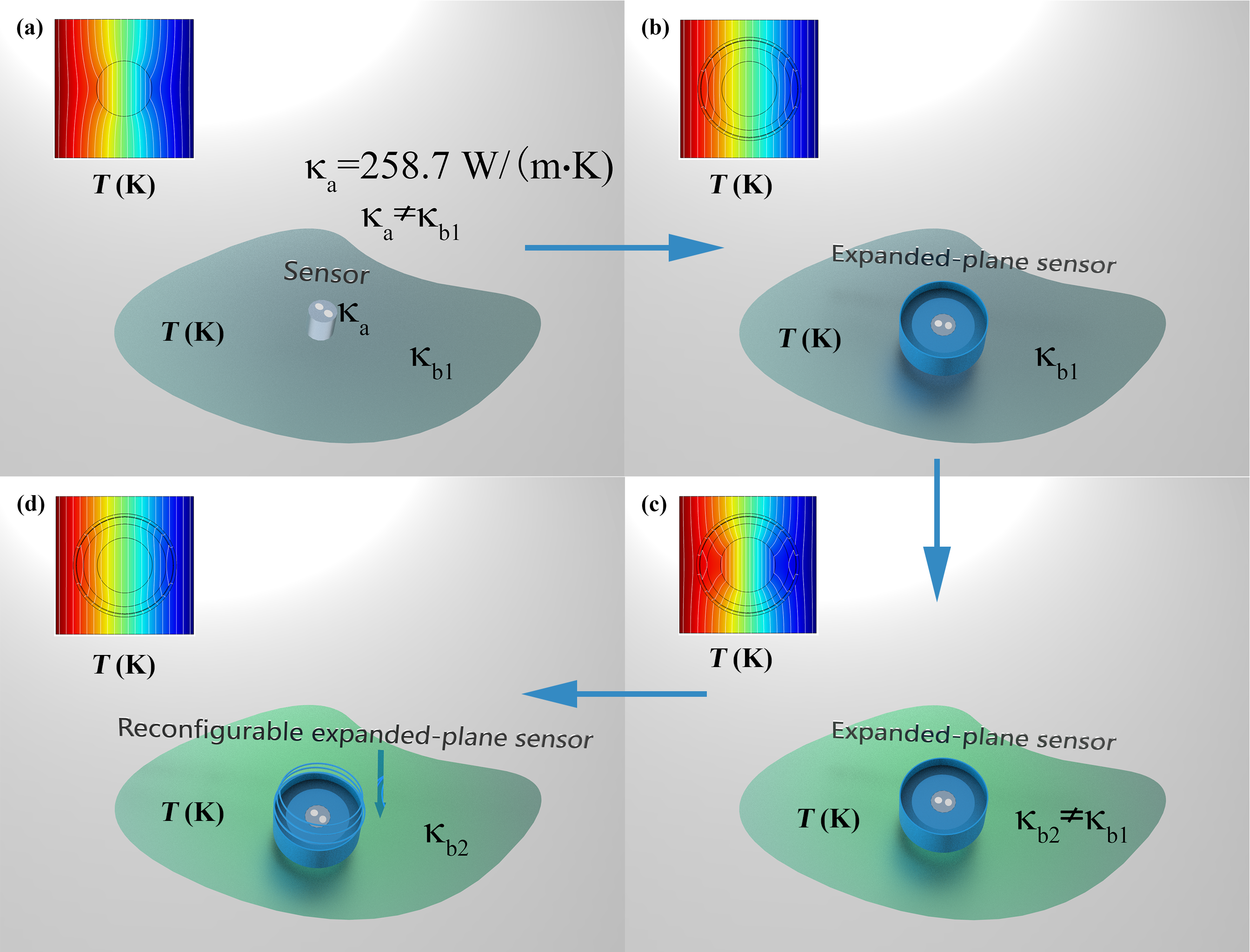}
  \caption{The figure illustrates: (a) a conventional sensor with a distorted temperature field in the background region ($\kappa_a \neq \kappa_{b1}$), (b) the expanded-plane sensor, which effectively eliminates the temperature distortions in the background region, and (c-d) the reconfigurable expanded-plane sensor's ability to maintain functionality even as the thermal conductivity of the background region changes from $\kappa_{b1}$ to $\kappa_{b2}$ by adjusting the height of the expanded plane.}
  \label{senshiyi}
\end{figure*}
\begin{figure*}[htp]
  \includegraphics[width=1.0\linewidth]{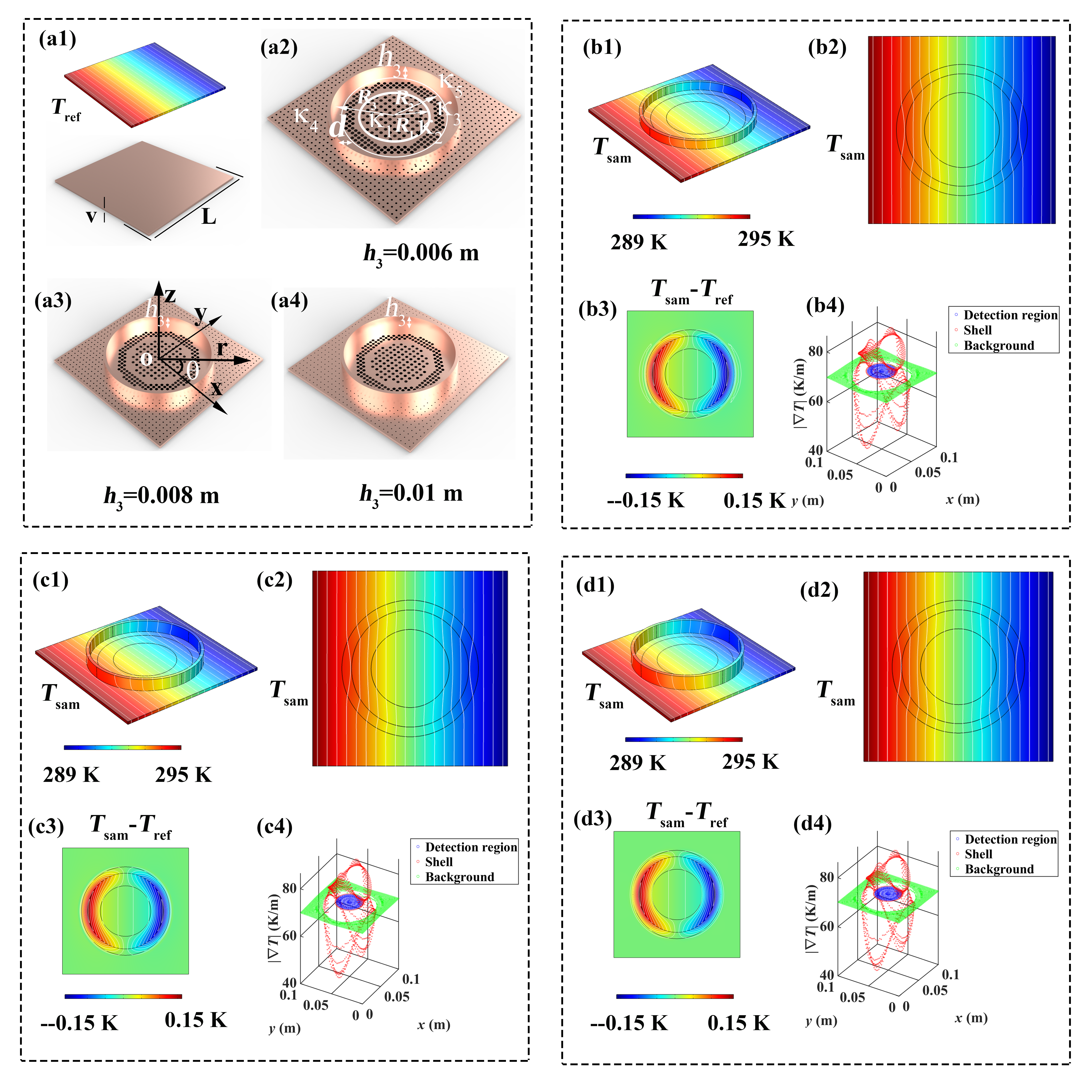}
  \caption{(a1) The structure of the reference and its temperature distribution are depicted. The simulation employs the following parameters: $L=0.01$ m, $R_1=0.002$ m, $R_2=0.003$ m, $R_3=0.035$ m, $v=0.002$ m, $d=0.002$ m, $\kappa_1=258.7$ W/m$\cdot$K, $\kappa_2=213.5$ W/m$\cdot$K, $\kappa_3=400$ W/m$\cdot$K, and $\kappa_5=400$ W/m$\cdot$K. The thermal conductivity of the background region are as follows:
  (a2) $\kappa_4=338.7$ W/m$\cdot$K;
  (a3) $\kappa_4=359.6$ W/m$\cdot$K;
  (a4) $\kappa_4=379.9$ W/m$\cdot$K.
  The corresponding height of expanded plane is $h_3=0.006$ m, $h_3=0.008$ m and $h_3=0.01$ m, respectively.
  (b1)-(b4) The simulation presents the temperature distribution, the temperature difference between structure in (a2) and the reference, and the temperature gradient for structure in (a2).
  (c1)-(c4) The corresponding simulation results are shown for structure in (a3).
  (d1)-(d4) The corresponding simulation results are shown for structure in (a4).}
  \label{senmoni}
\end{figure*}
\begin{figure*}[htp]
  \includegraphics[width=1.0\linewidth]{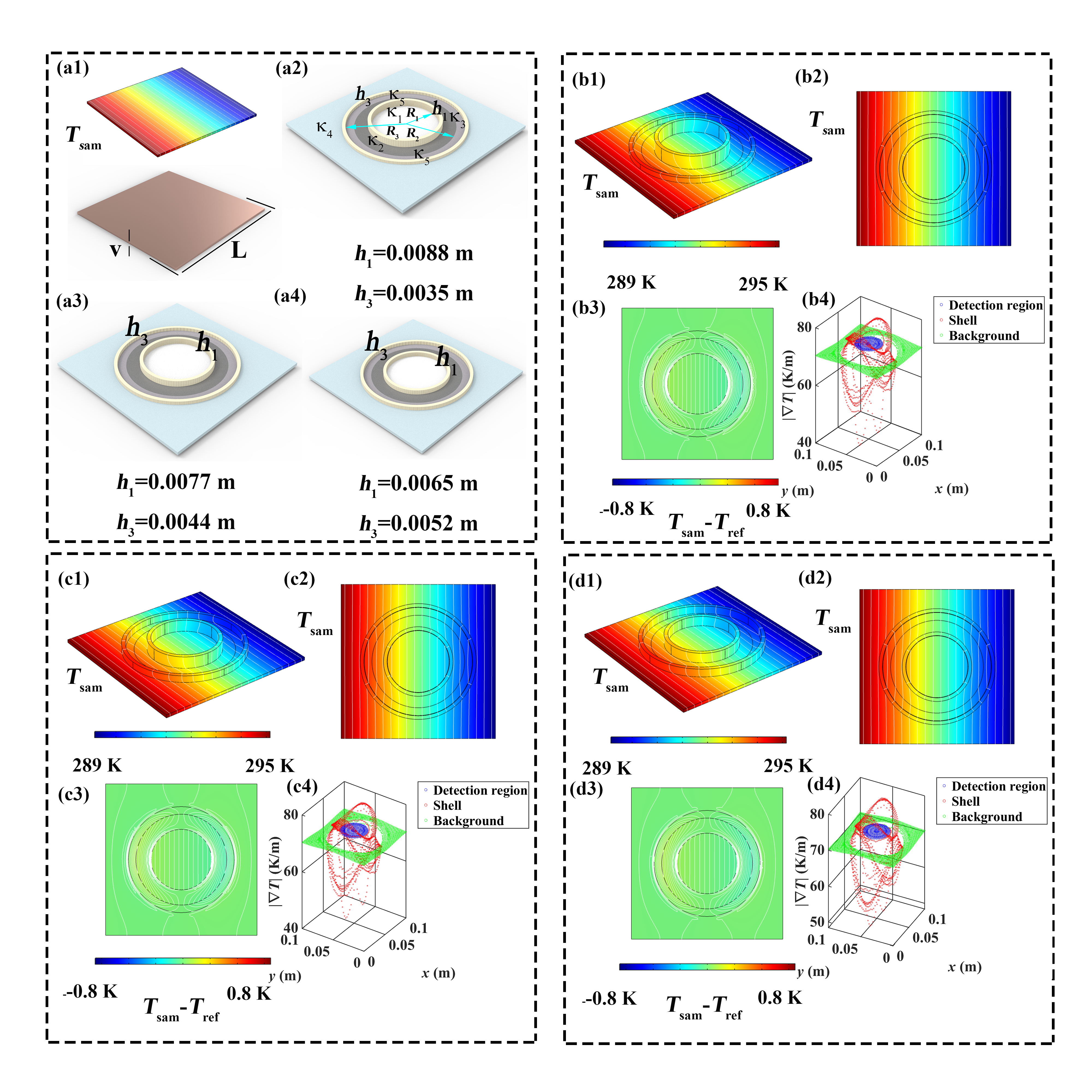}
  \caption{(a1) The figure illustrates the structure of the reference and its corresponding temperature distribution. Additionally, (a2)-(a4) present the structures of the MEP bilayer sensor with varying heights for the inner and outer expanded planes. The simulation employs the following parameters: $L=0.01$ m, $R_1=0.002$ m, $R_2=0.003$ m, $R_3=0.035$ m, $v=0.002$ m, $d=0.002$ m, $\kappa_1=190$ W/m$\cdot$K, $\kappa_2=213.5$ W/m$\cdot$K, $\kappa_3=400$ W/m$\cdot$K, $\kappa_5=400$ W/m$\cdot$K, $\kappa_6=400$ W/m$\cdot$K, and $h_1=0.0065$. The specific values for thermal conductivity of the background region are as follows:
  (a2) $\kappa_4=340.0$ W/m$\cdot$K;
  (a3) $\kappa_4=360.0$ W/m$\cdot$K;
  (a4) $\kappa_4=380.0$ W/m$\cdot$K.
  The corresponding height of outer expanded plane is $h_3=0.017$ m, $h_3=0.035$ m, and $h_3=0.0052$ m, respectively.
  (b1)-(b4) The simulation results depict the temperature distribution, the temperature difference between structure in (a2) and the reference, and the temperature gradient for structure in (a2).
  (c1)-(c4) The corresponding simulation results are presented for structure in (a3).
  (d1)-(d4) The corresponding simulation results are presented for structure in (a4). }
  \label{shuangsen}
\end{figure*}

\begin{figure*}[htp]
  \includegraphics[width=1.0\linewidth]{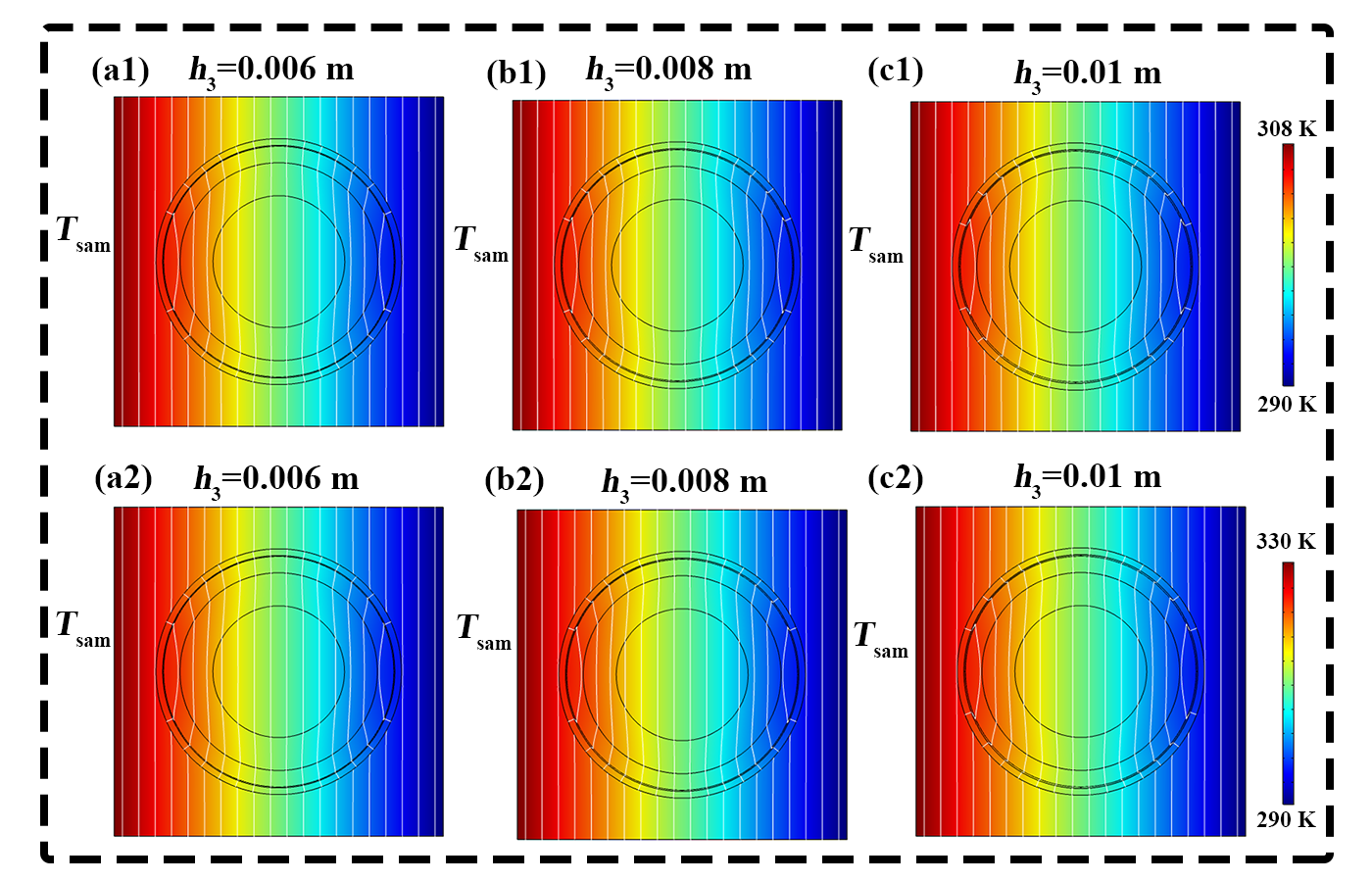}
  \caption{
  (a1)-(a2), (b1)-(b2), (c1)-(c2) The corresponding simulation results are presented for structure in Fig.~\ref{senmoni}(a2), (a3), (a4), with the temperature difference being $20$ K and $35$ K, respectively.}
  \label{fig5}
\end{figure*}
\begin{figure*}[htp]
  \includegraphics[width=1.0\linewidth]{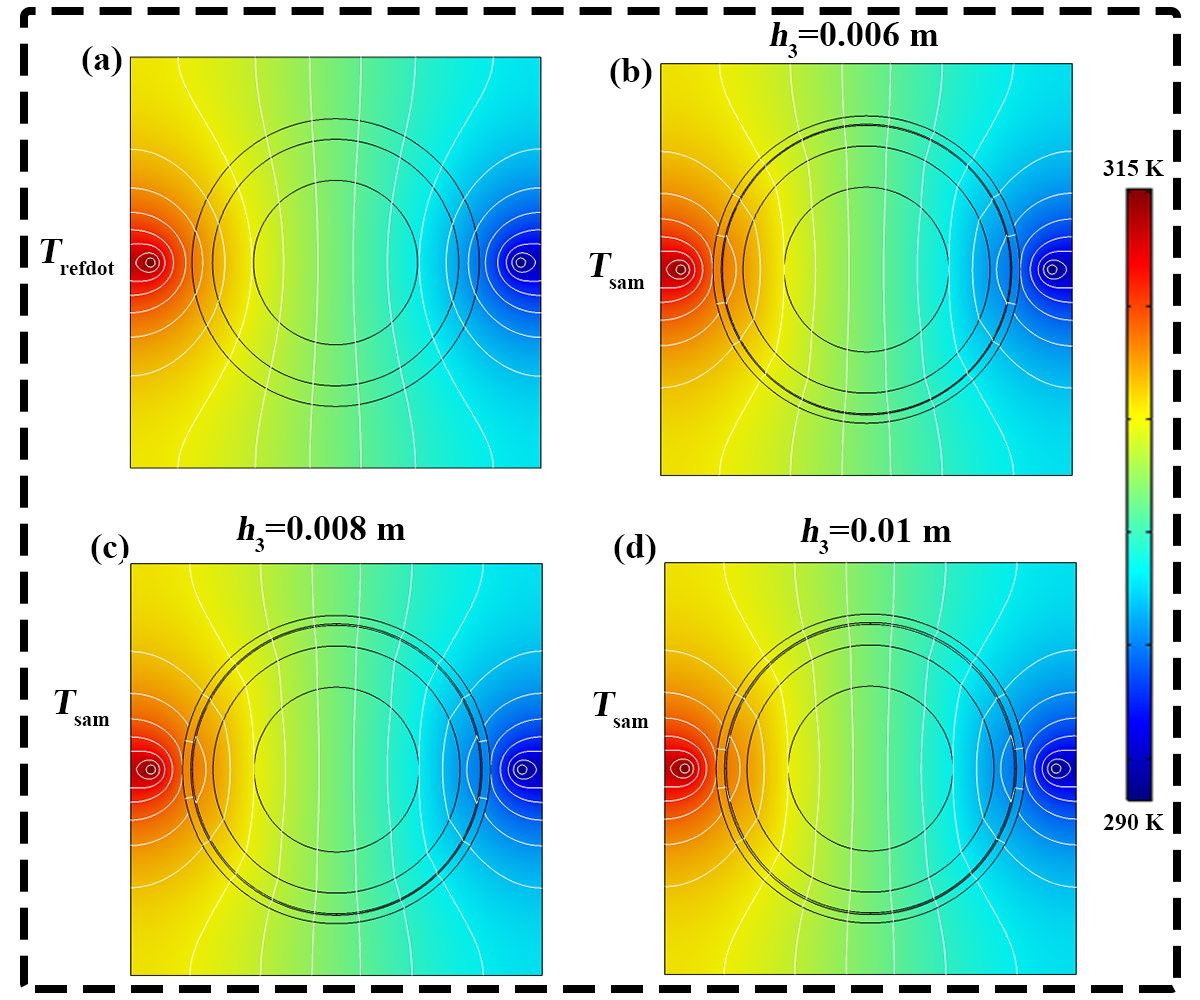}
  \caption{
   (a) The simulation results for the conventional two-dimensional structure with point hot source and cold source. The thermal conductivity of each region is the same.
    (b)-(d) The corresponding simulation results are presented for structure in Fig.~\ref{senmoni}(a2)-(a4) with point hot source and cold source.
   }
  \label{fig6}
\end{figure*}

\begin{figure*}[htp]
  \includegraphics[width=1.0\linewidth]{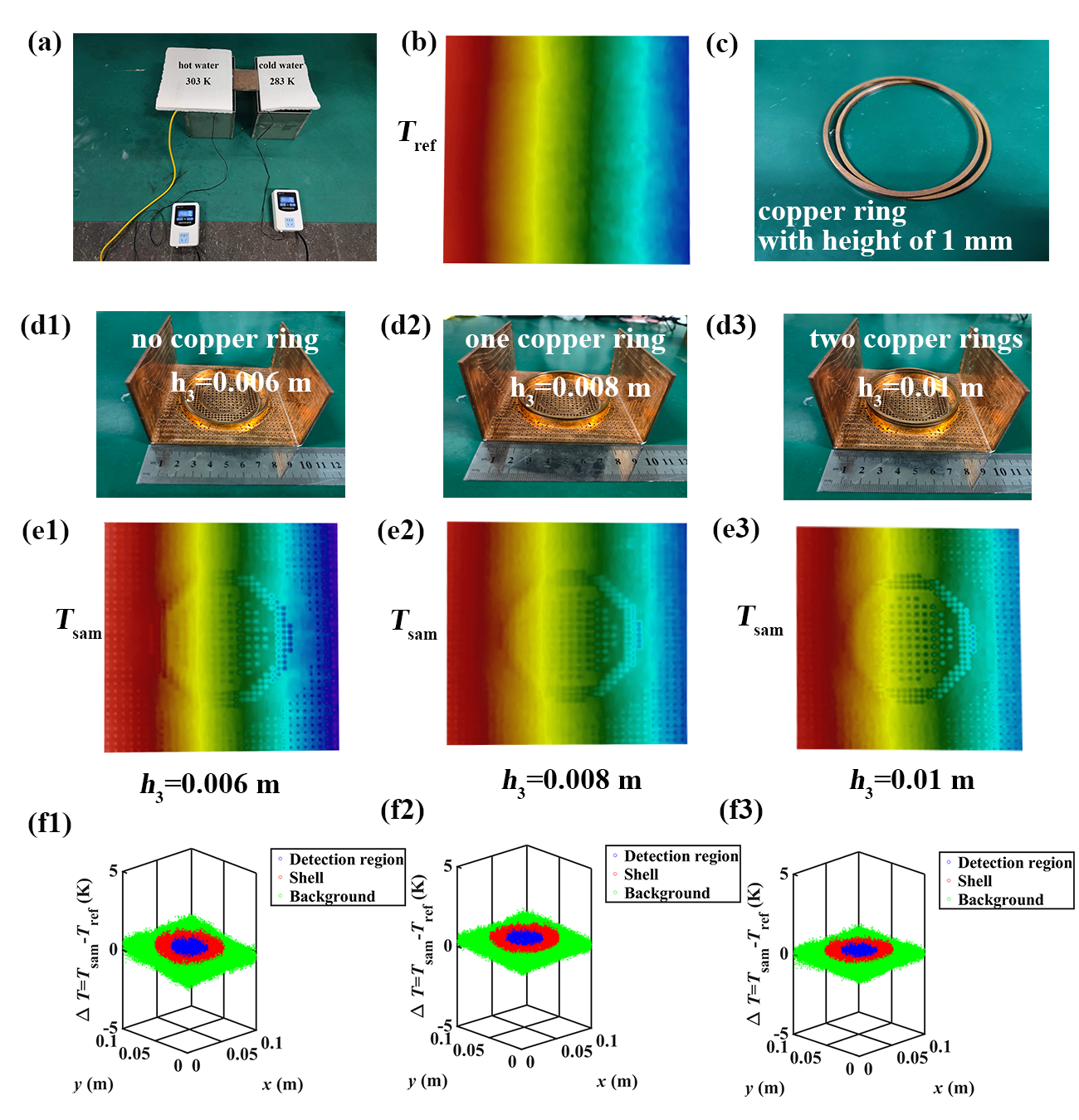}
  \caption{(a) The experimental setup is described.
  (b) The experimental results for the reference are presented.
  (c) Copper rings with a height of $2$ mm are utilized.
  (d1)-(d3) The structures of three samples are illustrated.
  (e1)-(e3) The experimental results for the three samples are shown.
  (f1)-(f3) The temperature difference between the samples and the reference is analyzed.}
  \label{senshiyan}
\end{figure*}

\begin{figure*}[htp]
  \includegraphics[width=1.0\linewidth]{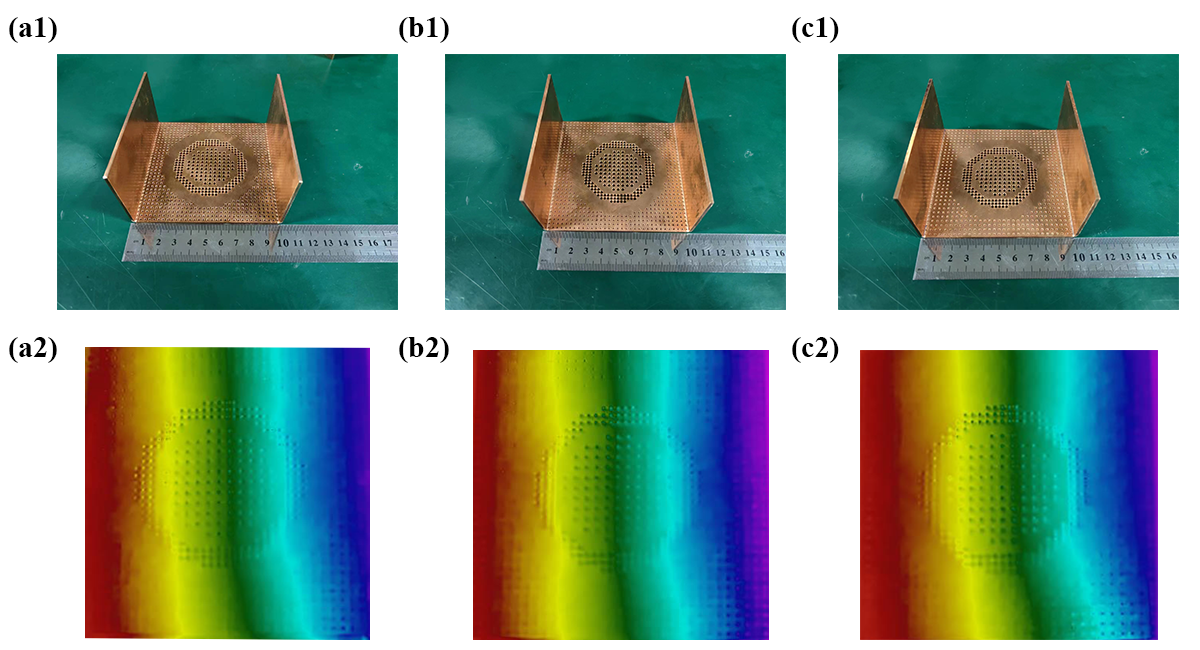}
  \caption{
    (a1)-(c1) The corresponding reference sample of Fig.~\ref{senmoni}(a2)-(a4) without expanded-plane.
    (a2)-(c2) The corresponding experimental results for (a1)-(c1).}
  \label{fig7}
\end{figure*}
\section{Problem Statement}
The structure of the SEP bilayer sensor is illustrated in Fig.\ref{senmoni}(a1)-(a2), where the substrate's dimensions are represented by $L$ (length), $L$ (width), and $v$ (thickness). The core, inner, and outer shell radii are denoted as $R_1$, $R_2$, and $R_3$, respectively. Additionally, the height and thickness of the outer expanded plane are specified as $h_3$ and $d$, respectively. Given that the thickness of both the substrate and the expanded plane is considered sufficiently small, vertical (along the $z$ direction) temperature variations on the substrate can be disregarded, and the temperature profile on the expanded plane only changes along the $z$ direction\cite{guo2022passive}. In order to achieve the desired sensor effect, it is crucial for the temperature profile in the background region to remain undistorted, with an equal temperature gradient between the core and the background region. {This is accomplished by solving the steady-state diffusion equation~\cite{sha2021robustly, zhang2023diffusion, xu2022diffusive}}, and the resulting temperature distribution for each region can be expressed as follows~\cite{xu2019bilayer}:
\begin{subequations}
  \begin{equation}
    T_1=Ar\cos\theta\ (r<R_1),
  \end{equation}
  \begin{equation}
    T_2=Br\cos\theta+\frac{C\cos\theta}{r}\ (R_1<r<R_2),
  \end{equation}
  \begin{equation}
    T_3=Dr\cos\theta+\frac{E\cos\theta}{r}\ (R_2<r<R_3),
  \end{equation}
  \begin{equation}
    T_4=Ar\cos\theta\ (r>R_3),
  \end{equation}
   \end{subequations}
   where $T_1$, $T_2$, $T_3$, and $T_4$ represent the temperature distribution of the core, inner shell, outer shell, and background region, respectively. The constants $A$, $B$, $C$, $D$, and $E$ depend on the boundary conditions. The distance between the field point and the origin is denoted by $r$, and $\theta$ represents the polar angle, as shown in Fig.~\ref{senmoni}(a3). We assume that the expanded plane's boundaries are adiabatic, except for the bottom boundary. Since the temperature is continuous at the boundary between the expanded plane and the substrate, we can obtain the temperature distribution on the expanded plane, $T_5$, as follows~\cite{guo2022passive}:
   \begin{equation}
     T_5=\frac{AR_3\cos\theta}{\cosh\left(\frac{h_3}{R_3}\right)}\cosh\left(\frac{h_3-z}{R_3}\right).
   \end{equation}
   Further considering the boundary conditions of temperature consistency and flux continuity between two adjacent regions, we can obtain the following system of equations:
   \begin{eqnarray}
    \label{test1}
      AR_3=DR_3+\frac{E}{R_3},\\
      \kappa_4A=\kappa_5A\tanh\left(\frac{h_1}{R_3}\right)+\kappa_3\left(D-\frac{E}{{R_3}^2}\right),\\
      DR_2+\frac{E}{R_2}=BR_2+\frac{C}{R_2},\\
      \kappa_3\left(D-\frac{E}{{R_2}^2}\right)=\kappa_2\left(B-\frac{C}{{R_2}^2}\right),\\
     BR_1+\frac{C}{R_1}=AR_1,\\
     \label{test2}
      \kappa_2\left(B-\frac{C}{R_1^2}\right)=\kappa_1A.
     \end{eqnarray}
     where $\kappa_1$, $\kappa_2$, $\kappa_3$, $\kappa_4$, and $\kappa_5$ represent the thermal conductivity of the core, inner shell, outer shell, background region, and expanded plane, respectively. Using Eq.~(\ref{test1})-Eq.~(\ref{test2}), we consider $B$, $C$, $D$, $E$, $h_3$, and $\kappa_3$ as the six variables, while other parameters are regarded as coefficients. We can obtain the expression of $h_3$ as a function of $R_1$, $R_2$, $R_3$, $\kappa_1$, $\kappa_2$, $\kappa_4$ and $\kappa_5$ (see Appendix). 

  \section{Finite-element simulations}
  To validate the theoretical analysis presented above, we conducted finite-element simulations using COMSOL MULTIPHYSICS. {Indeed, for solving the heat transfer problem, there are several alternative methods available, such as the Legendre wavelet collocation method~\cite{chaurasiya2023numerical, chaurasiya2023numerical1}, a hybrid numerical method based on Taylor-Galerkin and Legendre wavelets~\cite{chaurasiya2023taylor}, and the heat-balance integral method~\cite{chaurasiya2022study, chaurasiya2023temperature}. However, considering the advantages offered by the finite-element method, including geometry flexibility, mesh refinement capabilities, and its wide range of applications in the physical sciences, we chose it as our method of verification.}

  The simulation parameters were configured as follows: $L=0.01$ m, $R_1=0.02$ m, $R_2=0.03$ m, $R_3=0.035$ m, $\kappa_1=258.7$ W/(m$\cdot$K), $\kappa_2=213.5$ W/(m$\cdot$K), $\kappa_3=400$ W/(m$\cdot$K), and $\kappa_5=400$ W/(m$\cdot$K). Three different values of $\kappa_4$ were selected: $\kappa_4=338.7$ W/(m$\cdot$K), $\kappa_4=359.6$ W/(m$\cdot$K), and $\kappa_4=379.9$ W/(m$\cdot$K), each corresponding to different values of $h_3$ as follows: $h_3=0.006$ m, $h_3=0.008$ m, and $h_3=0.01$ m.
  
  The simulation results for $h_{3}=0.006$ m are depicted in Fig.\ref{senmoni}(b1)-(b4). These results clearly show that the background temperature profile remains undistorted. To provide a more vivid visualization of the sensor effect, we have plotted the temperature difference between the sample and reference in Fig.\ref{senmoni}(b3). Additionally, the temperature gradient in the core region matches that in the background, which is further emphasized by the temperature gradient distribution on the bottom surface, as illustrated in Fig.~\ref{senmoni}(b4).
  
  Furthermore, corresponding simulation results for $h_{3}=0.008$ m and $h_{3}=0.01$ m are presented in Fig.\ref{senmoni}(c1)-(c4) and Fig.\ref{senmoni}(d1)-(d4), respectively. These quantitative results highlight the excellent performance of the designed thermal sensor when operating in different background conditions.
  
  {To demonstrate the versatility of our design in accommodating various temperature fields, we adjusted the temperature difference between the linear heat sources. The simulation results confirm that our structure remains functional even when the temperature difference is altered (see Fig.~\ref{fig5}(a1)-(c2)). Additionally, we considered scenarios with point heat sources. By conducting finite-element simulations for the EP structures with different expanded planes, we demonstrated the effectiveness of the proposed structure. Therefore, our design exhibits applicability across a wide range of temperature fields.}

\section{Theory extension to the multiple expanded-planes case}
To extend the theoretical analysis presented above to the MEP case, we consider another expanded plane with a height of $h_1$ located in the core region (see Fig.~\ref{shuangsen}(a2)). The thickness of this plane can also be ignored. Then, the temperature distribution of the core expanded plane $T_6$ can be expressed as follows:
\begin{equation}
  T_6=\frac{AR_1\cos\theta}{\cosh\left(\frac{h_1}{R_1}\right)}\cosh\left(\frac{h_1-z}{R_1}\right).
\end{equation}
Accordingly, we replace Eq.~(\ref{test2}) by 
\begin{eqnarray}
  \kappa_2\left(B-\frac{C}{R_1^2}\right)=\kappa_1A+\kappa_6F\tanh\left(\frac{h_1}{R_1}\right),
 \end{eqnarray}
 where $\kappa_6$ represents the thermal conductivity of the inner expanded plane. In this case, we consider $B$, $C$, $D$, $E$, $h_1$, and $h_3$ as variables. By following a similar approach as before, we can obtain an expression for $h_1$ and $h_3$ as a function of the other parameters (see Appendix).
 
 \section{Finite-element simulations for the multiple expanded-planes case}

 In the MEP case, the values of $L$, $R_1$, $R_2$, $R_3$, $\kappa_3$, and $\kappa_5$ are set to the same values as in the SEP case. The values of $h_1$ and $\kappa_6$ are set as follows: $h_1=0.0065$ m, $\kappa_6=400$ W/(m$\cdot$K). Certain parameters are changed as the following values: $\kappa_1=190.0$ W/(m$\cdot$K), $\kappa_2=280.0$ W/(m$\cdot$K). To implement the thermal sensor, three sets of values are chosen for $\kappa_4$: $\kappa_4=340.0$ W/(m$\cdot$K); $\kappa_4=360.0$ W/(m$\cdot$K); $\kappa_4=380.0$ W/(m$\cdot$K). The corresponding values for $h_3$ are as follows: $h_3=0.0035$ m; $h_3=0.0044$ m; $h_3=0.0052$ m. Similar to the SEP case, the temperature distribution, temperature difference between the sample and the reference, and temperature gradient on the bottom surface are presented for the three cases. These results further demonstrate that the MEP structure ensures the excellent performance of a dynamic thermal sensor.

\section{Experiments}

To validate the theory of the reconfigurable expanded-plane thermal sensor, we conducted experiments in the context of the SEP case. Three copper samples were chosen due to their high thermal conductivity~\cite{yang2022transformation, zhuang2022nonlinear}, as illustrated in Fig.\ref{senshiyan}(d1)-(d3). The parameters for the first sample were defined as follows: $L=0.1$ m, $R_1=0.02$ m, $R_2=0.03$ m, $R_3=0.035$ m, $h_3=0.006$ m, $\kappa_1=258.7$ W/(m$\cdot$K), $\kappa_2=213.5$ W/(m$\cdot$K), $\kappa_3=400$ W/(m$\cdot$K), and $\kappa_4=338.7$ W/(m$\cdot$K). These parameters are consistent with those depicted in Fig.\ref{senmoni}(b1). To achieve the desired thermal conductivity for each region, we employed a specific number of holes following the principles of the effective medium theory (EMT)\cite{davis1977effective, tian2021thermal, bruggeman1935dielektrizitatskonstanten}. Assuming thermal conductivity values of $400$ W/(m$\cdot$K) for copper and $0.03$ W/(m$\cdot$K) for air, we determined area fractions of $21.5\%$, $30.4\%$, and $8.3\%$ for the core, inner shell, and background regions, respectively. The hole diameters were set at $0.002$ m, $0.001$ m, and $0.0006$ m, requiring drilling of $86$, $152$, and $602$ holes, respectively. This same methodology was applied to the second and third samples, with their parameters matching those of the simulated samples shown in Fig.\ref{senmoni}(c1) and (d1).

To realize the reconfigurable expanded-plane sensor, we prepared a series of copper rings each with a height of $2$ mm (Fig.~\ref{senshiyan}(c)) and adjusted the height of the expanded plane by adding or removing these rings. To compensate for changes in the thermal conductivity of the background region (in the case of the second and third samples), one ring and two rings were added to the expanded plane, respectively. {We filled thermal grease between the rings, allowing us to consider the thermal resistance of the expanded plane as negligible. Furthermore, as illustrated in Figs.\ref{senshiyan}(e1)-(e3), the experimental outcomes demonstrate enhanced thermal sensing performance. This observation corroborates the premise that the impact of the current contact thermal resistance is insignificant.}

The experimental setup is depicted in Fig.\ref{senshiyan}(a). One side of the sample was brought into contact with a hot source at a temperature of $303$ K, while the other side was connected to a cold source at $283$ K. Once the system reached a stable state, an infrared (IR) camera was employed to capture the surface of the sample from the top and record the temperature distribution. The resulting temperature distributions of the sample surfaces are presented in Fig.\ref{senshiyan}(e1)-(e3) for the three different samples. Additionally, the experimental result of the reference is shown in Fig.\ref{senshiyan}(b). To illustrate the thermal sensor effect, we present the temperature difference between the sample and the reference, as demonstrated in Fig.\ref{senshiyan}(f1)-(f3). These results validate the exceptional performance of the samples and affirm the accuracy of the thermal sensor theory.

{Furthermore, we conducted experiments for the reference sample without an expanded plane (see Fig.\ref{fig7}(a1)-(a3)). In this case, the thermal conductivity of each region matched that of the corresponding EP structure. The experimental results revealed temperature field distortion for structures lacking an expanded plane (see Fig.\ref{fig7}(a2)-(a3)). This further proves the effectiveness of our EP design.}

\section{Discussion and conclusion}
Expanding upon the single expanded-plane structure, we have successfully engineered a reconfigurable sensor. Our simulation and experimental results provide robust evidence of the device's ability to operate effectively across a wide range of background environments, achieved simply by adjusting the height of the expanded plane. Furthermore, we have extended our investigations to scenarios involving multiple expanded planes, and we have obtained consistent and reliable results. {In comparison to the conventional two-dimensional sensor structure, the expanded plane is an integral component of the proposed design. Nonetheless, we define the core region delineated by the radius $R_1$ as the operational area of the sensor, and the background region as the area external to the circle represented by the radius $R_3$, as depicted in Fig.\ref{senmoni}(a2)-(a4). To achieve the desired sensor effect, it is crucial to ensure that the thermal conductivity of each region matches appropriately~\cite{xu2020thermally}. According to conventional two-dimensional sensor theory, the thermal conductivity of the outer layer region should significantly exceed that of the background region. Although the thermal conductivity of the outer layer region for the expanded-plane structure is only $400$ W/(m$\cdot$K), the effective thermal conductivity of the outer layer can reach an ultra-high value since a portion of the heat flux is directed towards the expanded plane. This is why we can ensure that the temperature distribution remains undistorted using the EP structure.}

However, it is important to note that in this study, we have exclusively focused on the thermal conduction mechanism~\cite{xu2023black, Tan,yang2017full,xu2018thermal1,xu2020active}. It is imperative to acknowledge that the influences of thermal radiation~\cite{xie2023colored,xu2020transformation,xu2020transformation1,shen2016thermal1} and convection mechanisms~\cite{ju2023nonreciprocal,xu2021geometric,dai2018transient} should be considered for further investigation. To enhance the precision of our theoretical framework, it would be advisable to incorporate these variables into future research endeavors. {While we have exclusively presented results for a circular core region with a fixed size, it is worth highlighting that the proposed structure can also effectively operate with core regions of varying sizes and shapes~\cite{sha2022topology}. Moreover, we envision that the exploration of multiphysics effects could be facilitated by employing the expanded-plane configuration~\cite{yang2015invisible, lei2021temperature, lei2023spatiotemporal,zhuang2023multiple}. Furthermore, the proposed structure holds the potential for the design of other devices with innovative functionalities~\cite{gao2007magnetophoresis,dong2004dielectric,huang2003dielectrophoresis,ye2008non,liu2013statistical,huang2005magneto,qiu2015nonstraight}, such as concentrators.}

In conclusion, we have introduced an approach for implementing a reconfigurable sensor utilizing an expanded-plane structure. This study lays a robust foundation for the exploration of reconfigurability in multiphysics phenomena. Due to its straightforward fabrication process, our approach exhibits significant potential for expanding the application range of thermal metamaterials. It holds considerable value across various research domains, including thermal detection, energy management, material science, and thermal coding~\cite{hu2018binary, yang2023space}.


\section*{Declaration of Competing Interest}
The author declare that there are no conflicts of interest.
\printcredits

\section*{Acknowledgment}
We gratefully acknowledge funding from the National Natural Science Foundation of China (Grants No. 12035004 and No. 12320101004) and the Innovation Program of Shanghai Municipal Education Commission (Grant No. 2023ZKZD06). We also would like to express our gratitude to Dr. Gaole Dai of Nantong University for his valuable and insightful suggestions that have helped us improve our research work.

\section*{Appendix}
In the case of single expanded plane, the relationship between the height of the expanded plane, denoted as $h_3$, and other parameters can be expressed as follows:
\begin{align}
  h_3&=R_3\cdot \arctanh\left[(\kappa_4 - s\cdot q/t)/\kappa_5\right].
\end{align}
Here, the value of $s$ is given by $s=m/n-p$. The specific expressions for $m$, $n$, $p$, $q$, and $t$ are as follows:
\begin{align}
  &m=50\cdot \left(\kappa_1\cdot R_2^2 - \kappa_1\cdot R_1^2
  + \kappa_2\cdot R_1^2+ \kappa_2\cdot R_2^2\nonumber \right.\\ 
   &\left.- 2\cdot \kappa_2\cdot R_3^2\right),  \\
  &n=\kappa_2\cdot \left(R_2^2 - R_3^2\right), \\
  &p=50\cdot \left(R_1^2 - R_2^2\right)\cdot \left(\kappa_1 - \kappa_2\right)/\left(\kappa_2\cdot \left(R_2^2- R_3^2\right)\right), \\
  &q=- \kappa_2^2\cdot R_1^2\cdot R_2^2 + \kappa_2^2\cdot R_1^2\cdot R_3^2+\kappa_2^2\cdot R_2^4 \nonumber \\
  &- \kappa_2^2\cdot R_2^2\cdot R_3^2 + \kappa_1\cdot \kappa_2\cdot R_1^2\cdot R_2^2 + \kappa_1\cdot \kappa_2\cdot R_2^4\nonumber \\
  &- \kappa_1\cdot \kappa_2\cdot R_1^2\cdot R_3^2 -\kappa_1\cdot \kappa_2\cdot R_2^2\cdot R_3^2, \\
  &t=100\cdot \left(\kappa_1\cdot R_2^4 + \kappa_2\cdot R_2^4- \kappa_1\cdot R_1^2\cdot R_2^2\nonumber \right.\\
  &- \kappa_1\cdot R_1^2\cdot R_3^2+ \kappa_2\cdot R_1^2\cdot R_2^2+\kappa_1\cdot R_2^2\cdot R_3^2\nonumber \\
  &\left.+ \kappa_2\cdot R_1^2\cdot R_3^2- 3\cdot \kappa_2\cdot R_2^2\cdot R_3^2\right).
\end{align}
In the multiple expanded-planes case, we obtain the similar expressions:
\begin{align}
  h_1&=R_1\cdot \arctanh \left\{\left[\kappa_1\cdot \kappa_3\cdot R_2^4 - \kappa_1\cdot \kappa_2\cdot R_2^4\nonumber \right.\right.\\
   &- \kappa_2^2\cdot R_2^4 + \kappa_2\cdot \kappa_3\cdot R_2^4 + \kappa_2^2\cdot R_1^2\cdot R_2^2\nonumber \\ 
   &- \kappa_2^2\cdot R_1^2\cdot R_3^2 + \kappa_2^2\cdot R_2^2\cdot R_3^2 - \kappa_1\cdot \kappa_2\cdot R_1^2\cdot R_2^2\nonumber \\
  &+ \kappa_1\cdot \kappa_2\cdot R_1^2\cdot R_3^2 - \kappa_1\cdot \kappa_3\cdot R_1^2\cdot R_2^2 \nonumber \\
  &+ \kappa_1\cdot \kappa_2\cdot R_2^2\cdot R_3^2 - \kappa_1\cdot \kappa_3\cdot R_1^2\cdot R_3^2\nonumber \\
  &+ \kappa_2\cdot \kappa_3\cdot R_1^2\cdot R_2^2 + \kappa_1\cdot \kappa_3\cdot R_2^2\cdot R_3^2\nonumber \\
  &\left.+ \kappa_2\cdot \kappa_3\cdot R_1^2\cdot R_3^2 - 3\cdot \kappa_2\cdot \kappa_3\cdot R_2^2\cdot R_3^2 \right]/ \left[\kappa_6\nonumber \right.\\
  & \times \left(\kappa_2\cdot R_2^4- \kappa_3\cdot R_2^4 + \kappa_2\cdot R_1^2\cdot R_2^2 \nonumber \right.\\
  &- \kappa_2\cdot R_1^2\cdot R_3^2 + \kappa_3\cdot R_1^2\cdot R_2^2- \kappa_2\cdot R_2^2\cdot R_3^2\nonumber \\ 
  &\left.\left.\left.+ \kappa_3\cdot R_1^2\cdot R_3^2 - \kappa_3\cdot R_2^2\cdot R_3^2 \right) \right]\right\}.\\
  h_3&=R_3\cdot \arctanh \left\{\left[\kappa_2\cdot \kappa_3\cdot R_2^4 - \kappa_3^2\cdot R_2^4 \nonumber \right.\right.\\
  &+ \kappa_2\cdot \kappa_4\cdot R_2^4- \kappa_3\cdot \kappa_4\cdot R_2^4 + \kappa_3^2\cdot R_1^2\cdot R_2^2 \nonumber \\
   &- \kappa_3^2\cdot R_1^2\cdot R_3^2+ \kappa_3^2\cdot R_2^2\cdot R_3^2 - 3\cdot \kappa_2\cdot \kappa_3\cdot R_1^2 \nonumber \\
    &\times R_2^2+ \kappa_2\cdot \kappa_3\cdot R_1^2\cdot R_3^2+ \kappa_2\cdot \kappa_4\cdot R_1^2\cdot R_2^2  \nonumber \\
     &+ \kappa_2\cdot \kappa_3\cdot R_2^2\cdot R_3^2- \kappa_2\cdot \kappa_4\cdot R_1^2\cdot R_3^2 \nonumber \\
      &+ \kappa_3\cdot \kappa_4\cdot R_1^2\cdot R_2^2 - \kappa_2\cdot \kappa_4\cdot R_2^2\cdot R_3^2 \nonumber \\
       &\left.+ \kappa_3\cdot \kappa_4\cdot R_1^2\cdot R_3^2- \kappa_3\cdot \kappa_4\cdot R_2^2\cdot R_3^2 \right]/\left[\kappa_5 \nonumber \right.\\
       &\times \left(\kappa_2\cdot R_2^4 - \kappa_3\cdot R_2^4+ \kappa_2\cdot R_1^2\cdot R_2^2- \kappa_2\cdot R_1^2\nonumber \right.\\
        &\times R_3^2+ \kappa_3\cdot R_1^2\cdot R_2^2 - \kappa_2\cdot R_2^2\cdot R_3^2 + \kappa_3\cdot R_1^2\nonumber \\
        &\left.\left.\left. \times R_3^2- \kappa_3\cdot R_2^2\cdot R_3^2 \right) \right]\right\}.
\end{align}

\bibliographystyle{elsarticle-num}
\balance




\end{document}